\documentclass{article}
\usepackage{authblk}

\usepackage[T1]{fontenc}

\usepackage[utf8]{inputenc}
\usepackage{geometry}
\usepackage{changepage}

\usepackage{comment}

\usepackage{textcomp}
\usepackage{float}
\usepackage{amsmath, amssymb}
\usepackage{braket}
\usepackage[english]{babel}
\usepackage{dirtytalk}
\usepackage[separate-uncertainty=true]{siunitx}
\usepackage{graphicx}
\usepackage[labelfont=bf]{caption}
\usepackage{subcaption}
\usepackage{mathtools}
\usepackage{physics}
\usepackage{amssymb}
\usepackage{amsmath}
\usepackage{mathtools}
\usepackage{xfrac}
\usepackage{array}
\usepackage{relsize}
\usepackage{tabu}
\usepackage{bbold}
\usepackage{tensor}

\usepackage[font=small]{caption}
\usepackage{amsfonts}
\usepackage{booktabs}
\usepackage{siunitx}
\usepackage{appendix}
\usepackage{mathrsfs}
\usepackage{amsthm}
\usepackage{indentfirst}
\usepackage{soul} 
\setstcolor{orange}
\usepackage{xcolor}
\usepackage{hyperref} 
\usepackage{xcolor} 
\setstcolor{cyan}

\numberwithin{equation}{section}
\numberwithin{figure}{section}

\newcommand\christoffel[3]{\Gamma\indices{^#1_#2_#3}}

\usepackage{systeme}
\usepackage{tensor}

\title{Asymptotically Conically Minkowskian spacetimes from self-gravitating dust}
\author{Matteo Fontana\footnote{mfontana11@uninsubria.it}, Federico Scali\footnote{fscali@uninsubria.it}, Sergio Luigi Cacciatori\footnote{sergio.cacciatori@uninsubria.it}\\ 
Disat, Universit\`a dell'Insubria, via Valleggio 11, 22100 Como, Italy \\  INFN, Sezione di Milano, via Celoria 16, 20133 Milano, Italy\\}
\date{}

\begin{document}

\maketitle

\begin{abstract}

In this work we investigate some non-Newtonian effects in exact solutions of the Einstein equations, which describe stationary and axisymmetric configurations of self-gravitating dust. A distinctive feature of these solutions is the potential presence of conical singularities along the rotation axis, manifesting as angular deficits. While such singularities can be removed by imposing suitable boundary conditions along the axis, asymptotically far away from it the geometry becomes locally flat, in the sense that the Riemann tensor vanishes, but globally, instead of reducing to Minkowski space, it takes the form of a cone. We refer to these spacetimes as Asymptotically Conically Minkowskian (ACM).
We show that such conical structure can originate some interesting effects as seen by asymptotic local observers. These include modifications to the gravitational lensing and the misinterpretation of the vacuum state of a scalar field as a distribution of scalar particles.

\end{abstract}

\section{Introduction}

Galaxy dynamics has long been understood within the framework of Newtonian mechanics, since the low velocities and weak gravitational fields of galaxies seem to suggest that General Relativity (GR) would play only a minor role. However, a significant problem arises when this traditional approach fails to explain the non-Keplerian velocity profiles observed in galaxies, particularly the nearly flat rotation curves that persist far from their centers, suggesting the presence of more mass than is directly observed \cite{Zwicky:1933gu,Zwicky:1937zza,Cacciatori2024}.
This situation could indicate either a limitation in the theory of gravity being used or a need for additional matter beyond what is observed. One prevailing approach has been to keep the Newtonian framework by introducing dark matter (DM), a hypothetical form of matter that interacts primarily through gravity with ordinary baryonic matter. Dark matter not only helps explaining rotation curves but also plays a crucial role in the $\Lambda$CDM cosmological model, facilitating structure formation \cite{Primack:1997av} and providing insights into various astrophysical phenomena that appear to require additional mass. These include the virial estimates of galaxies within galaxy clusters \cite{lopez}, gravitational lensing effects \cite{Massey_2010}, thermodynamic emissions of X-rays in galaxy clusters \cite{Castillo-Morales:2003mdk}, the Bullet Cluster \cite{Clowe:2003tk,Markevitch:2003at}, and the peaks observed in the Cosmic Microwave Background \cite{Dvorkin:2022bsc}. Numerous hypotheses have been proposed regarding the exact composition of DM, ranging from Weakly Interacting Massive Particles to Axion-Like Particles and Massive Compact Halo Objects \cite{Cacciatori2024}. Despite extensive efforts to detect and characterize DM, its exact nature remains a subject of debate. Alongside the challenges posed by dark energy, this quest to comprehend the dark sector invites a reexamination of the $\Lambda$CDM model and its implications for our understanding of the universe.
Importantly, all the observations challenging the Newtonian approach have gravitational nature; therefore, it could be possible to explain them by keeping the matter sector with the known particles while modifying the theory of gravity. This insight led to the formulation of Modified Newtonian Dynamics (MOND) \cite{Milgrom:1983ca, Milgrom:1983pn, Milgrom:1983zz}, suggesting that Newton's law of gravitation would break down in regions of very low acceleration, like galaxy edges, where gravity would weaken more slowly than predicted by the inverse-square law, thus explaining the flat rotation curves of galaxies without the need of DM. Another alternative is provided by MOdified gravity theories (MOGs) \cite{CANTATA:2021asi}, which aim to extend or alter general relativity to explain cosmic acceleration and galactic dynamics without invoking dark matter or dark energy \cite{CacciatoriE2024}. 
However, since the currently accepted theory of gravity, i.e., General Relativity (GR), already departs significantly from Newtonian theory, it is worthwhile to explore the potential of fully general relativistic models of galaxies. \\
The fact that Newtonian theory, intended as the weak field, small velocities limit of General Relativity, is not able to account for the velocity curves of disk galaxies without DM is generally regarded as the failure of either Einstein's theory or the Standard Model of elementary particles, or even both. However, unlike Newtonian gravity, the Einstein Equations (EE) are non-linear and the gravitational field itself carries energy and momentum. Moreover, solutions to the EE possess more degrees of freedom than Newtonian gravity, which is described by a single scalar potential. 
When applied to galaxies, these features can significantly impact the theoretical explanation of the known phenomenology. A paradigmatic consequence is the
frame dragging effect displayed by the off-diagonal components of the metric, which are in general of the same order of the Newtonian potential \cite{Crosta:2020,Beordo:2024,Astesiano:2021ren}, see also \cite{Astesiano:2022ghr,Ludwig:2024mgo}. This observation suggests that, although Newtonian mechanics is often used as a weak-field, low-velocity limit of general relativity, it may be inadequate for a global description of spatially extended rotating systems like galaxies: general relativistic effects become increasingly important on large scales, where the non-linearities of GR and the additional degrees of freedom it introduces can lead to observable phenomena which are not captured by the Newtonian approximation. Therefore, it makes sense to seek exact solutions of the Einstein equations for modeling extended objects and explore the conditions under which, even at small velocities and energy densities, GR provides additional information that Newtonian gravity is missing.
In this context, after the pioneering work carried out in previous studies, the first viable fully general relativistic model for a disk galaxy was presented by Balasin and Grumiller (BG) \cite{BG}, eliminating some unphysical behavior of the earlier proposed solutions \cite{Cooperstock2005,Cooperstock:2006dt,Cooperstock:2006ti,Cooperstock:2007sc}. This model describes a disc galaxy in terms of a stationary and axially symmetric solution of the EE coupled to the energy-momentum tensor of rigidly rotating dust, along the equatorial plane and far away from the rotation center. In particular, it allows to determine a simple expression for the velocity profile possibly compatible with some observed curves. Their result suggest that general relativistic effects may contribute to the explanation of a fraction of the dark energy content.\\
The BG velocity profile has been tested on the Milky Way from $5$ up to $19\, kpc$ from the galactic center by fitting the data of a subsample made of $719 143$
stars (including $241 918$ OBA stars, $475 520$ RGB giants, and $1705$ Cepheides) selected from a sample of nearly $33$ million stars from the GaiaDR3 catalogue \cite{Crosta:2020,Beordo:2024}. The results led to the conclusion that the BG velocity profile is statistically indistinguishable from its state-of-the-art DM analogues.  At this point it is worth to mention that there is quite a confusion in the scientific literature about the extent of the analysis performed in \cite{Crosta:2020,Beordo:2024}. While the authors use the BG velocity profile, they do not start from the BG model for the fitting, but from the general Lewis--Papapetrou--Weyl (LPW) structure \cite{Stephani:2003tm} without selecting a particular analytic solution. Therefore, these analyses are not affected by the defects of the BG solution which is a very particular case of the LPW family.  \\
These general relativistic dust models cannot provide a description of an entire galaxy but only of a portion where the dust approximation is acceptable. However, they already demonstrate that GR can play a significant role in understanding the required amount of dark matter. Therefore, it is important to understand to what extent the nonlinearity of general relativity can lead to a discrepancy with Newtonian gravity on large scales, under non-relativistic conditions, and it is natural to wonder if other tests can be performed to prove the viability of such models, focusing on their non-Newtonian effects. \\
An important characteristic of general relativistic models is the conical asymptotic behavior observed far from the rotation axis, manifesting as angular defects in the spacetime geometry. This globally non Minkowskian behavior offers observable effects that are not taken into account in the Newtonian limit; hence they can be considered a signature of the exact general relativistic description of extended rotating objects. As a consequence, studying the observable effects of locally flat and conical structures in axially symmetric backgrounds acquires importance, since it can provide indications upon the need of a general relativistic description of galaxies. In the following we will avoid referring to these solutions as conical defects, as the spacetime near the axis is modified by the presence of matter and boundary conditions can always be chosen such that no conical singularity is present on the axis. Instead we will refer to them as ``Asymptotically conical Minkowskian'' (ACM) spacetimes. We also remark that conical singularities are not in general physical singularities, as witnessed by a vast literature on cosmic strings, see e.g. \cite{Vilenkin}. \\
After an introduction to the stationary, axisymmetric models of self-gravitating dust and the emergence of ACMs we discuss three observables which rely on the global topological nature of such defects and could prove the ACM structure.
A direct approach involves analyzing the holonomy group by considering a global observer that takes a closed loop lacing up the symmetry axis and parallel transporting a vector along the way; this vector will test a nontrivial holonomy after closing the loop, confirming a conical structure, while a local observer confined in a limited convex region not intersecting the axis would perceive the geometry as globally flat. 
Another method involves the examination of the light deflection caused by spacetime geometry, which allows a straightforward comparison between general relativity predictions and observations. 
Finally, another possibility leverages the differences in the propagation of a quantized scalar field perceived by a global and a local observer. Those are particularly evident in the respective definitions of the vacuum states and could be exploited in ideal scattering experiments to expose the conical geometry.

\

The paper is organized as follows. In Sec. \ref{Stationary axisymmetric spacetimes} we introduce the general metric of a stationary and axisymmetric spacetime coupled to dust, and show how the conical geometry may arise in both singular and regular solutions. In Sec. \ref{topological defects} we explore in more details the presence of the asymptotic conical geometry without conical singularities. In particular, in Sections \ref{holonomy method} and \ref{lensing} we investigate the possible implications of the conical topology on the parallel transport of a test vector along a closed loop around the symmetry axis, and the gravitational lensing. Section \ref{quantum} is devoted to the study of quantum effects: a true global vacuum state for a scalar field results to be interpreted by a local observer (not aware of the conical structure) as a distribution of scalar particles.
Section \ref{conclusions} is devoted to conclusions, while some mathematical details are demanded to Appendices \ref{Appendix A} and \ref{Appendix B}.

\section{Stationary axisymmetric self-gravitating dust}
\label{Stationary axisymmetric spacetimes}
We consider a disc of self-gravitating dust, which is modeled as a stationary and axisymmetric solution of the Einstein Equations (EE). The latter is characterized by two Killing vectors: a timelike vector $\xi$, and a spacelike vector $\zeta$, whose Killing product vanishes. With the additional assumption that the spacetime admits 2-spaces orthogonal to the Killing vectors, as discussed in \cite{Stephani:2003tm} the metric in cylindrical coordinates $(t,r,z,\varphi)$ takes the general Lewis--Papapetrou--Weyl (LPW) form 
\begin{equation}
\label{General SA metric dust}
    ds^{2}=e^{-2U}\left[e^{2k}(dr^{2}+dz^{2})+W^{2}d\varphi^{2}\right]-e^{2U}(dt+Ad\varphi)^{2},
\end{equation}
where $U(r,z)$ is related to the Newtonian potential; $A(r,z)$ is responsible for frame dragging; and $k(r,z)$ is a conformal factor on the 2-spaces orthogonal to the orbits of the Killing vectors.
The stress-energy tensor of the dust is 
\begin{equation}
    T_{\mu\nu}=\rho u_{\mu}u_{\nu},
\end{equation}
where $\rho$ is the energy density and $u^{\mu}$ the four-velocity of the dust particles. Given the symmetries of the spacetime, the latter reads
\begin{equation}
\label{dust velocity}
    u=\frac{1}{\sqrt{-H(r,z)}}(\partial_{t}+\Omega(r,z)\partial_{\varphi}),
\end{equation}
where $\Omega=\frac{d\varphi}{dt}$ is the dust angular velocity. The function $H(r,z)$ is determined by the normalization condition $u^{\mu}u_{\mu}=-1$, yielding
\begin{equation}
\label{expression for H}
    H(r,z)=e^{-2U}W^{2}\Omega^{2}-e^{2U}(1+A\Omega)^{2}.
\end{equation}
The metric \eqref{General SA metric dust} can be further simplified by setting $W(r,z)=r$ thanks to the structure of the EE.

\subsection{Rigidly rotating self-gravitating dust}
\label{BG model}
In this section we specialize to the case of rigidly rotating dust as described by the BG model \cite{BG}. \\
We start by remarking that this model cannot be used to describe a whole galaxy and is still problematic for globally modeling self-gravitating dust. For such a purpose, a more general LPW metric would be necessary. However, it is the simplest possible solution already capturing the effects we are interested in, which in the most general case would change only in magnitude but not in form
 
The BG solution is obtained by assuming the angular velocity $\Omega$ of the system to be a constant, so that in the comoving system one can set $\Omega=0$. Due to the conservation of the stress-energy tensor the four-velocity of dust is geodetic; hence, the Killing vector $\xi$ is affine geodetic. Consequently, the function $U(r,z)$ appearing in the metric \eqref{General SA metric dust} becomes a constant that can be reabsorbed in a redefinition of the time variable\footnote{This also implies $H=-1$.}. The metric then simplifies to 
\begin{equation}
\label{rigidly rotating dust metric}
    ds^{2}=-(dt-Nd\varphi)^{2}+r^{2}d\varphi^{2}+e^{\mu}(dr^{2}+dz^{2}).
\end{equation}
The solution to the EE is then obtained by means of a spectral separation of variables \cite{BG}
\begin{align}
 N(r,z)=\int_0^\infty d\lambda \mathcal{R}(r,\lambda)\mathcal{Z}(z,\lambda).
\end{align}
This approach leads to
\begin{equation}
\label{eq: generic solution for N in and out the galaxy plane}
    N(r,z)=\frac{r^{2}}{2}\int_{0}^{\infty}dx\ C(x)\Big([(z+x)^{2}+r^{2}]^{-\frac{3}{2}}+[(z-x)^{2}+r^{2}]^{-\frac{3}{2}}\Big),
\end{equation}
which is expressed in terms of an arbitrary spectral function $C(x)$. 
In the spirit of \cite{BG}, we ask the following less restrictive conditions:

\begin{itemize}
    \item $v(r,0)\propto r$ for small values of $r$; 
    \item $v(r,0)\propto \frac{1}{r}$ asymptotically, well outside the disc,
\end{itemize}
where $v$ is the rotation velocity as measured by ZAMOs (Zero Angular Momentum Observers).
The most simple spectral function that provides a velocity profile satisfying these conditions is 
\begin{equation}
\label{BG spectral density}
    C(x)=(x-r_{0})v_{0}[\theta(x-r_{0})-\theta(x-R)]+(R-r_{0})\theta(x-R),
\end{equation}
where $\theta(x)$ is the Heaviside step function; $v_{0}$ is a constant velocity (relative to the speed of light); $r_{0}$ is an internal radius; and $R$ is the radius of the disc of dust. 
With this choice Eq.\ \eqref{eq: generic solution for N in and out the galaxy plane} becomes
\begin{equation}
\label{eq: simplified expression of N for the particular choice of spectral density}
    N(r,z)=v_{0}(R-r_{0})+\frac{v_{0}}{2}\left[\sqrt{(z+r_{0})^{2}+r^{2}}+\sqrt{(z-r_{0})^{2}+r^{2}}-\sqrt{(z+R)^{2}+r^{2}}-\sqrt{(z-R)^{2}+r^{2}}\right].
\end{equation}
We will consider this solution in a thin disc $|z|<r_0$, so that $N(0,z)=0$. \\
The next step is the determination of $\mu$, the other free function of the metric \ref{rigidly rotating dust metric}. To simplify the analysis, we restrict our attention to the galactic plane $z=0$, yielding
\begin{align}
\label{eq: explicit expression for mu}
    \mu(r,0)=&-\frac{v_0^2}{2}\left[\frac{1}{2}\ln(r^2+r_0^2)+\frac{1}{2}\ln(r^2+R^2) - \ln(r_0^2+R^2+2r^2+2\sqrt{r^2+r_0^2}\sqrt{r^2+R^2})\right] \notag \\
    &+ \frac{v_0^2}{2}\ln(\frac{Rr_0}{(R+r_0)^2})+ \mu_0. 
\end{align}
where $\mu_0=\mu(0,0)$ is an integration constant. Notice that to avoid a conical singularity on the rotation axis we must impose $\mu_0=0$. 

Some remarks about the physical viability of the BG solution are in order. First, it is important to acknowledge its limited range of applicability to the dust model, which requires sufficient proximity to the galactic plane. Specifically, for small r and large $z$ it may happen that $r^{2}\leq N^{2}$, causing the $\varphi$ coordinate to become timelike. In such cases the orbits of the Killing vector $\partial_{\varphi}$ (circles of constant $r$ and $z$) become closed timelike curves; therefore, to avoid this issue we will mostly restrict the analysis to $|z|<r_0$, which suffices for the purposes of the present article.\\ 
Nevertheless, let us now shortly discuss the general case.

\subsection{Differentially rotating self-gravitating dust}
\label{differentially rotating dust}
To provide a more realistic description of self-gravitating dust, it is necessary to relax the assumption of rigid rotation. In this context, we consider the ($\eta,H)$ models \cite{Stephani:2003tm,Astesiano:2021ren}, which represent a class of stationary and axisymmetric solutions of the EE coupled to dust, where differential rotation is allowed ($\Omega_{a}\neq 0, a=r,z)$. 
The name comes from the fact that the metric, as well as the dust density and four-velocity, are parametrized 
by an arbitrary function $\eta(r,z)$ and a related negative function $H(\eta)$. More precisely, 
\begin{align}
    g_{tt}=&\frac{(H-\eta\Omega)^{2}-r^{2}\Omega^{2}}{H},\\
    g_{t\varphi}=&\frac{\eta^{2}-r^{2}}{-H}\Omega+\eta,\\
    g_{\varphi\varphi}=&\frac{r^{2}-\eta^{2}}{-H},\\
    g_{rr}=& g_{zz}= e^{\mu(r,z)},\\
    \mu_{r}=&\frac{1}{2r}[(g_{tt})_{r}(g_{\varphi\varphi})_{r}-(g_{tt})_{z}(g_{\varphi\varphi})_{z}-((g_{t\varphi})_{r})^{2}+((g_{t\varphi})_{z})^{2}],\\
    \mu_{z}=&\frac{1}{2r}[(g_{tt})_{z}(g_{\varphi\varphi})_{r}+(g_{tt})_{r}(g_{\varphi\varphi})_{z}-2(g_{t\varphi})_{r}(g_{t\varphi})_{z}],
\end{align}
where
\begin{equation}
\label{equation for omega}
    \Omega=\frac{1}{2}\int H'\frac{d\eta}{\eta}
\end{equation}
gives the angular velocity of the dust with respect to the $(t,\varphi)$ coordinates.\\
Correspondingly, the energy density of the dust is
\begin{equation}
\label{energy density}
    8\pi G\rho=\frac{\eta^{2}r^{-2}(2-\eta l)^{2}-r^{2}l^{2}}{4g_{rr}}\frac{\eta_{r}^{2}+\eta_{z}^{2}}{\eta^{2}},
\end{equation}
with
\begin{equation}
    l=\frac{1}{H}\frac{dH}{d\eta}.
\end{equation}
Finally, notice that $\eta$ and $H$ cannot be chosen completely independently, in the sense that for an arbitrary choice of $H(\eta)$, one can introduce the auxiliary function
\begin{equation}
\label{def F}
    \mathcal{F}=2\eta+r^{2}\int \frac{H'}{H}\frac{d\eta}{\eta}-\int\frac{H'}{H}\eta d\eta,
\end{equation}
which is constrained by the EE to satisfy
\begin{equation}
\label{armonicity of F}
    \mathcal{F}_{rr}-\frac{1}{r}\mathcal{F}_{r}+\mathcal{F}_{zz}=0.
\end{equation}
Replacing \eqref{def F} into \eqref{armonicity of F}, one gets an equation for $\eta(r,z)$, 
\begin{equation}
\label{equation for eta}
    \left(\eta_{ r r}-\frac{1}{r} \eta_{ r}+\eta_{ z z}\right)(2-\eta \ell)+\left(\eta_{ r}^2-\eta_{ z}^2\right)\left(\eta \ell^{\prime}-\ell\right)\left(1+\frac{r^2}{\eta^2}\right)+r^2 \frac{\ell}{\eta}\left(\eta_{ r r}+\frac{3}{r} \eta_{ r}+\eta_{ z z}\right)=0.
\end{equation}
With respect to the ZAMOs, the velocity of the dust is given by 
\begin{equation}
    v_{Z}(r,z)=\frac{\eta(r,z)}{r}.
\end{equation}
We also remark that the form $\mu_r dr+\mu_z dz$ is closed and so locally integrable. 
However, finding exact solutions in this general situation is much more difficult than in the rigid case. Nevertheless, we can make the following considerations. For simplicity, let us assume symmetry w.r.t. the galactic plane to which we restrict our considerations, so that all first order derivatives in $z$ vanish. When $r\to0$, $\eta(r,0)$ is expected to go to zero as $\eta(r,0)\sim r^2$, while $H$ will tend to a constant, $H\sim H_0<0$. Regularity in $r=0$ also requires that next to leading order in $H$ grows at least like $r^2$. However, to have $\Omega\sim \Omega_0$ in $r\sim 0$, we assume $H\sim H_0+a r^3$ there. From these considerations we see that when $r\sim 0$, the metric $d\sigma^2$ of the surface $t=t_0$, $z=0$, is nearly
\begin{align}
 d\sigma^2\sim e^{\mu_0} dr^2+\frac {r^2}{-H_0} d\varphi^2,
\end{align}
so that we have to impose $e^{-\mu_0}=-H_0$ to avoid a conical singularity. Notice that we have not normalized $-H_0$ to 1 since we reserve such normalization to the limit $r\to\infty$ where we assume the metric to become static and flat.  \\
Therefore, when $r\to\infty$ still on the galactic plane, we ask $H\to-1$, $\eta \sim 1/r$, and $r^2\Omega\to0$. Having fixed these conditions we are not left with further freedom and so
\begin{align}
 \mu\sim \mu_\infty= \mu_0+\int_0^\infty \frac{1}{2r}[(g_{tt})_{r}(g_{\varphi\varphi})_{r}-((g_{t\varphi})_{r})^{2}] dr.
\end{align}
We then see that the metric at infinity on the same surface is
\begin{align}
 d\sigma^2\sim e^{\mu_\infty} dr^2+r^2 d\varphi^2.
\end{align}
Unless a special coincidence occurs such that $\mu_\infty=0$, imposing the absence of conical singularities may lead to a conical geometry far away from the symmetry axis. Notice that not only the surface but also the whole asymptotic metric has vanishing curvature; therefore, local observers restricted to a finite angular sector of spacetime can easily redefine coordinates and make the metric Minkowskian. Of course, this is not feasible globally because of the conical topology. 

This conical geometry can cause deviations in the observable properties of the self-gravitating dust that are not captured by Newtonian physics, making them a distinct hallmark of the general relativistic approach. It is worth to mention that this kind of behavior should not be expected just for the case of dust, but most probably for the majority of the extended systems which are axially symmetric and stationary, as disc galaxies may be. In the next sections we explore how this nontrivial topology can be probed through various observable tests, allowing us to assess whether the geometry of spacetime aligns with the predictions of general relativity or adheres more closely to the expectations of the Newtonian approximation.

\section{ACM spacetimes from self-gravitating dust}
\label{topological defects}
The possibility of an asymptotically conical geometry around the rotation axis is a purely non-Newtonian effect of general relativity that can be revealed by the analysis of both classical and quantum observables. At this point, we want to remark that our viewpoint in the present article is substantially different from the one in \cite{Galoppo:2023jsf}, where the presence of conical singularities has been used to restrict the physical applicability of the van Stockum-Bonner class of solutions. On the opposite, we are arguing that axially symmetric configurations of extended matter systems free of conical singularities may present an asymptotic conical geometry that is only locally flat, and represents a potential source of new interesting physics, distinct from Newtonian predictions, rather than a defect.
Moreover, even when present along a symmetry axis, the angular defects do not manifest as physical singularities and can play a significant role in cosmology \cite{Vilenkin}. \\
Asymptotically conical spacetimes are well approximated by flat globally conical manifolds far away from the conical singularity of the latter. 
A prototypical example of such a manifold is described by the following line element, often used to model the spacetime of a cosmic string:
\begin{equation}
    ds^{2}=-dt^{2}+dr^{2}+\frac{r^{2}}{\mu^{2}}d\theta^{2}+dz^{2},
\end{equation}
where $t\in\mathbb{R}$, $r\in[0,\infty)$, $\theta\in[0,2\pi)$, $z\in\mathbb{R}$ and $\mu\in\mathbb{R}$. Introducing the variable $\Tilde{\theta}=\frac{\theta}{\mu}$ the metric takes the form 
\begin{equation}
    ds^{2}=-dt^{2}+dr^{2}+r^{2}d\Tilde{\theta}^{2}+dz^{2}.
\end{equation}
Locally this metric is isomorphic to the Minkowski one but not globally, since the period of the angular variable is now $\frac{2\pi}{\mu}$ and we assume $\mu\neq1$, thus introducing an angular defect. Therefore, the spacetime contains a conical singularity that can only be detected by completing a revolution around it.

The structures we have in mind give rise to spacetimes that are well approximated by this geometry when $r$ is large enough. 
This could generate a problem concerning the astrophysical measurements performed by local observers who are not aware of the angular defect. Indeed, think of galaxies as potentially carrying such a structure, endowed with local observers that can only employ Earth based telescopes or satellites. When global observables come into play, a local observer is likely to interpret differently the experimental data compared to, say, an observer that has taken a closed loop around the axis and thus knows of the angular deficit (or excess). This will be a recurrent theme in the following. 
\\
We can now proceed by considering the explicit example of the Balasin-Grumiller model, where close to the rotation axis $(r\ll r_{0})$ the metric components can be expanded as 
\begin{equation}
    N(r)\underset{r\ll r_{0}}{\simeq}\frac{v_{0}}{2}\left(\frac{1}{r_{0}}-\frac{1}{R}\right)r^{2},
\end{equation}
\begin{equation}
    e^{\mu(r)}\underset{r\ll r_{0}}{\simeq}e^{\mu_{0}}.
\end{equation}
Specializing to the 2-dimensional surfaces $\{t=const.,z=const.\}$ the metric \eqref{rigidly rotating dust metric} at the lowest order reads
\begin{equation}
\label{BG metric close to axis}
    ds^{2}\underset{r\ll r_{0}}{\simeq} r^{2}d\phi^{2}+e^{\mu_{0}}dr^{2}.
\end{equation}
If we want to avoid conical singularities we have to put $\mu_0=0$. From  Eq.\ \eqref{eq: explicit expression for mu}, we then see that at infinity\footnote{Notice that for the BG model at $r\to\infty$ the time part of the metric enters as $(dt-N_\infty d\phi)^2$ with $N_\infty$. We will see quickly how this affects the holonomy.}
\begin{align}
\label{BG metric far from axis}
 ds^{2}\underset{r\gg R}{\simeq} r^{2}d\phi^{2}+e^{\mu_{\infty}}dr^{2}
\end{align}
where
\begin{align}
 \mu_\infty= -\frac {v_0^2}2 \ln \frac {(R+r_0)^2}{4r_0R},
\end{align}
which is strictly negative for $r_0<R$.\\
Performing the change of variable $\Tilde{r}=re^{\frac{\mu_{\infty}}{2}}$, Eq.\ \eqref{BG metric far from axis} becomes 
\begin{align}
    ds^{2}\underset{r\gg R}{\simeq}& \Tilde{r}^{2} \frac {d\phi^{2}}{\lambda^2}+d\Tilde{r}^{2},\label{ridotta}\\
    \lambda^2 =& \left(\frac {(r_{0}+R)^{2}}{4r_{0}R}\right)^{-\frac{v_{0}^{2}}{2}}<1, 
\end{align}
from which we finally recognize a flat conical geometry on the 2-surfaces $\{t=const.,z=const.\}$ at large distances from the symmetry axis, with an angular excess. \\

\subsection{Walking around the cone: the holonomy method}
\label{holonomy method}
As discussed in the previous section, detecting the conical geometry requires global methods that involve exploring an extended region of spacetime.
A first and very direct approach involves the study of the nontrivial holonomy group of the metric space \cite{Oliveira1996}, by means of the parallel transport of a test vector in a closed loop around the symmetry axis. Since this procedure must be performed in the asymptotic region, the exact metric can be replaced with its asymptotic expression in order to compute the asymptotic holonomy. Let us briefly review how this works.
\\
Consider an $m-$dimensional Lorentzian manifold $\mathcal{M}$ equipped with an affine connection $\nabla$. At every point $p$ of $\mathcal{M}$, the connection naturally defines a group of linear transformations from the tangent space $T_p\mathcal{M}$ into itself, $T_p\mathcal{M} \rightarrow T_p\mathcal{M}$. Indeed, consider a closed path $\gamma$ starting at $p$, $\{\gamma(t): 0\leq t\leq 1, \gamma(0) = \gamma(1) = p\}$. Take a vector $v_0$ in $T_p\mathcal{M}$ and consider its parallel transport from $p$ to $p$ along $\gamma$. The result is a transported vector $v_t$ again belonging to $T_p\mathcal{M}$ for $t=1$. Therefore, the connection $\nabla$ and the loop $\gamma$ together define a transformation $h_p(\gamma): T_p\mathcal{M} \rightarrow T_p\mathcal{M}$ which is linear and invertible. The set of all transformations $Hol_p = \{h_p(\gamma)\}$ at varying $\gamma$ is a group, the group operation being the composition of paths, and it is called the \textit{holonomy group} of the point $p$ \cite{Nakahara2003}.
It is also immediate to realize that the subset of all transformations generated by the contractible loops (i.e., the loops which can be continuously shrunk to a point) is a subgroup of $Hol_p$, known as the \textit{restricted holonomy group} $Hol_p^0$.\\
To be more concrete, let us consider a closed loop $\gamma: t\mapsto \gamma(t)$ starting from $p$ and a vector $v\in T_p\mathcal{M}$. Its parallel transport $v_t:=v(t)$ along $\gamma$ is governed by the equation $\dot \gamma^\mu \nabla_\mu v^\nu=0$, from which we see that 
\begin{align}
 h_p(\gamma)v=v+\Delta v, 
\end{align}
with
\begin{align}
 \Delta v^\nu=\oint_\gamma \partial_\mu v^\nu dx^\mu=-\oint_\gamma \Gamma^\nu_{\mu\rho} v^\rho dx^\mu.
\end{align}
If the manifold is smooth and $\gamma$ is contractible we can take a surface $S$ with area element $d\Sigma^{\mu\nu}$ and apply Stokes' theorem to write
\begin{align}
 \Delta v^\nu=\frac 12 \int_S R^\nu_{\ \rho\mu\sigma} v^\rho d\Sigma^{\mu\sigma}.
\end{align}
In the limit when the curve is contracted to a point, the holonomy group element thus goes to the identity. In this sense, one can see the $so(1,m-1)$ valued\footnote{To be more precise, if $g(p)$ is the metric in $p$ this form takes values in $so(g(p))$.} two form in $p$ 
\begin{align}
 {\omega_p}^\nu_{\ \rho} =\frac 12 R^\nu_{\ \rho\mu\sigma}(p) d\Sigma^{\mu\sigma}(p)
\end{align}
as the infinitesimal generator form of the holonomy group in $p$.\\
The situation changes if the spacetime is singular at $p$. In this case, the matrix $h_p$ can be defined by considering closed paths encircling $p$ (or the singular locus containing $p$) and then gradually shrinking them to the point. Notice that, because $p$ is singular, the holonomy matrix fails to become the identity even in this limit. Also, it is important to remark that the holonomy groups associated to points connected by a path are conjugate and so isomorphic; the group element responsible for the conjugation being determined by the parallel transport along the connecting path. In particular, when a loop is shrunk to $p$ its base point draws a path ending in $p$; hence, the holonomy groups associated to the points of the resulting curve are isomorphic.

With all this in mind, the idea is fairly simple. We have to consider large loops lying far away in the asymptotically conical region and encircling once the symmetry axis. To compute their associated holonomy we can then replace the true metric with the conical one and notice that each loop of the kind we are considering can be connected to the axis through a straight line (or any curve), along which the Riemann tensor is always zero. The holonomy we are looking for is thus completely determined by that along the symmetry axis of the prototypic conic manifold, which can be determined by restricting on the plane $t=0$, $z=0$. In particular, we can choose the path $t=0, z=0, r=\bar r, \theta\in[0,2\pi]$. 

After rescaling $r=\tilde r/\lambda$, the relevant metric is \eqref{ridotta}, with nonvanishing Christoffel symbols
\begin{align}
 \Gamma^\phi_{\tilde r \phi}=\frac 1{\tilde r}, \qquad \Gamma^r_{\phi\phi}=-\frac {\tilde r}{\lambda^2}.
\end{align}
Starting from the vector $(v^\phi, v^{\tilde r})$, the parallel transport equations read
\begin{align}
\begin{pmatrix}
 dv^\phi \\ dv^{\tilde r}
\end{pmatrix}
=
\begin{pmatrix}
 0 & -\frac 1{\tilde r} \\ \frac {\tilde r}{\lambda^2} & 0
\end{pmatrix}
\begin{pmatrix}
 v^\phi \\ v^{\tilde r}
\end{pmatrix} d\phi,
\end{align}
which have solution
\begin{align}
\begin{pmatrix}
 v^\phi(\phi) \\ v^{\tilde r}(\phi)
\end{pmatrix}= 
\begin{pmatrix}
 \cos \frac {\phi}{\lambda} & -\frac {\lambda}{\tilde r} \sin \frac {\phi}{\lambda} \\
 \frac {\tilde r}{\lambda} \sin \frac {\phi}{\lambda} & \cos \frac {\phi}{\lambda}
\end{pmatrix}
\begin{pmatrix}
 v^\phi \\ v^{\tilde r}
\end{pmatrix}.
\end{align}
The holonomy matrix is thus 
\begin{align}
 M=\begin{pmatrix}
 \cos \frac {2\pi}{\lambda} & -\frac {\lambda}{\tilde r} \sin \frac {2\pi}{\lambda} \\
 \frac {\tilde r}{\lambda} \sin \frac {2\pi}{\lambda} & \cos \frac {2\pi}{\lambda}
\end{pmatrix}=
\begin{pmatrix}
 \frac {\lambda}{\tilde r} & 0 \\
 0 & 1
\end{pmatrix}
\begin{pmatrix}
 \cos \frac {2\pi}{\lambda} & - \sin \frac {2\pi}{\lambda} \\
 \sin \frac {2\pi}{\lambda} & \cos \frac {2\pi}{\lambda}
\end{pmatrix}
\begin{pmatrix}
 \frac {\tilde r}{\lambda} & 0 \\
 0 & 1
\end{pmatrix},
\end{align}
which is in the congruence class of a rotation of an angle $2\pi/\lambda$. It becomes trivial if and only if $\lambda=1$, which corresponds to a globally Minkowskian structure instead of a conical one.

This result can be understood by considering the topological structure of the spacetime as perceived by two distinct classes of observers: a global and a local one. The former is characterized by the ability to travel all around the galaxy, for example following the geodesic of a dust particle, and therefore to perform the proposed experiment of parallel transporting a vector along a closed loop around the rotation axis. By comparing the initial vector with the one obtained after a complete loop they can read out the holonomy matrix and detect the presence of a conical geometry. In contrast, the local observer is confined in a limited angular region of the galaxy and cannot take a full trip around the axis. 
As a consequence, when presented with the result of a round trip the skeptical, local observer would always change the coordinates in a way that eliminates the conformal factor and reduces the holonomy matrix to the identity; in agreement with their preconception that the spacelike sections of spacetime are just planes.
\\
The simple calculation we have just done has to be slightly modified in the BG case, since the function $\eta(r,z)$ does not go to zero at infinity but tends to a constant $N_\infty$. At very large $r$ the metric contains the term $(dt-N_\infty d\phi)^2$ and, since $\phi$ is periodic, one may expect an holonomy term involving also $v^t$ components (mixed with $v^\phi$). To see this, let us use a more precise expression for the asymptotic $BG$ metric.
In Appendix A, the Christoffel symbols are expanded for $r \to \infty$ on the plane $z=0$ up to order $\frac{1}{r^2}$. The transport equations can again be solved along the circular orbit as above and up to order $\frac{1}{r^2}$ they give 
\begin{equation}
\begin{aligned}
    &v^r(2\pi) \simeq \cos(2\pi e^{-\mu/2}) v_0^r + \frac{V_0}{4r^2}e^{-\mu/2}(R^2-r_0^2)\sin(2\pi e^{-\mu/2})v_0^t +\\
    &\qquad\qquad -e^{-\mu/2}(\frac{V_0}{2r^2}(R-r_0)(R^2-r_0^2)-r)\sin(2\pi e^{-\mu/2})v_0^\phi,\\
    &v^t(2\pi) = e^{-\mu/2}\left(\frac{V_0}{r}(R-r_0) - \frac{V_0}{2r^2}(R^2 - r_0^2 + (R-r_0)^2) \right)\sin(2\pi e^{-\mu/2})v_0^r + v_0^t + \\ &\qquad\qquad + (-V_0(R-r_0) + \frac{V_0}{2r}(R^2-r_0^2 + (R-r_0)^2))(1-\cos(2\pi e^{-\mu/2}))v_0^\phi,\\
    &v^\phi(2\pi) = -\frac{1}{r}e^{\frac \mu2} \sin(2\pi e^{-\mu/2})v_0^r + \cos(2\pi e^{-\mu/2})v_0^\phi,\\
    &v^z(2\pi) = v_0^z,     
\end{aligned}
\end{equation}
where we have already evaluated the system at $\phi=2\pi$, and $\lambda=e^{\frac {\mu_{\infty}}{2}}$. Therefore, the holonomy matrix in the limit $r\gg R$ is (here the coordinates are ordered as $(t,r,\phi,z)$)
\begin{equation}
\lim\limits_{r\to\infty}h_\infty(\gamma)= 
\begin{bmatrix}
    1&0& -V_0(R-r_0)(1-\cos(2\pi e^{-\mu_{\infty/2}})) & 0 
    \vspace{3mm} \\
    0&\cos(2\pi e^{-\mu_{\infty}/2})&re^{-\mu_{\infty}/2}\sin(2\pi e^{-\mu_{\infty}/2})& 0
    \vspace{3mm}\\
    0&-\frac{1}{r}e^{-\mu_{\infty}/2} \sin(2\pi e^{-\mu_{\infty}/2})& \cos(2\pi e^{-\mu_{\infty}/2})& 0
    \vspace{3mm}\\
    0&0&0&1
\end{bmatrix}.
\end{equation}
This coincides with the above calculation, with the correction due to the $N_\infty= -V_0(R-r_0)$ term.\\
We can then conclude that an asymptotic, global observer taking a loop around the axis of a general relativistic dust model is able to realize the conical geometry of the $\{t=const.,z=const.\}$ surfaces by using the holonomy method; that is, by parallel transporting a test vector along the loop and confronting it with the initial one. 

We stress that, if a post-Newtonian approximation is adopted to describe the gravitational fields of the rotating dust, no conical geometry arises. Moreover, there is no reason to think that such a difference would be restricted to the case of dust; in fact, we should expect it for any extended rotating object. For example, this locally flat geometry could be expected outside disc galaxies. 
The holonomy test discussed above suggests an ideal experiment. Take a satellite orbiting the galaxy in the galactic plane, equipped with a gyroscope which gets Fermi-Walker transported along the satellite's world line. By comparing the gyroscope's orientation at a starting point and after one loop, it is in principle possible to measure the holonomy matrix and reveal the nontrivial topology of the galactic plane. This would potentially demonstrate the need for general relativistic galaxy models and their non-Newtonian features. Of course, such an experiment could be set up, at most, for a self-gravitating disc of dust with a few light-years of diameter, so that the satellite could complete a round trip in a reasonable time. Hence, the holonomy experiment is not a realistic possibility to test the outer topology of a galaxy. \\
A more effective way to investigate the global geometry surrounding a galaxy is the analysis of gravitational lensing, as observed in the behavior of pairs of light rays that trace a closed path around the galaxy.

\subsection{Null rays in the Balasin-Grumiller model}
\label{lensing}
The gravitational lensing in BG has been already studied in \cite{Galoppo2022}; however, in that case the aim was to show the excessively huge time delay generated by the rigid dragging at large scales, and the angular excess in the specific model was considered perturbatively too small to influence the results of the paper.
Instead, here we are interested in analyzing the possible influence of the conical geometry and check its measurability; therefore, we will repeat here the calculation using a different computational framework.\\
Because of the spacetime curvature, the path of light from distant stars or other sources gets deflected as it passes near massive objects. In the presence of conical singularities this deflection is further modified, offering a unique way to probe the angular deficits associated with these singularities. \\
Let us start by specifying the conserved quantities associated to the symmetries of the spacetime under consideration. These are given by the Killing vector fields $\xi=\partial_t$ and $\zeta=\partial_\phi$,
\begin{equation}
\label{energy}
    E=-g_{\mu\nu}\xi^{\mu}\frac{dx^{\nu}}{d\lambda}=-g_{tt}\frac{dt}{d\lambda}-g_{t\varphi}\frac{d\varphi}{d\lambda},
\end{equation}
\begin{equation}
\label{angular momentum}
    J=g_{\mu\nu}\zeta^{\mu}\frac{d x^{\nu}}{d\lambda}=g_{t\varphi}\frac{dt}{d\lambda}+g_{\phi\phi} \frac{d\varphi}{d\lambda}. 
\end{equation}
Furthermore, we can consider the interval 
\begin{equation}
\label{epsilon}
    \frac{ds^{2}}{d\lambda^{2}}=-g_{\mu\nu}\frac{dx^{\mu}}{d\lambda}\frac{dx^{\nu}}{d\lambda}=-g_{tt}\left(\frac{dt}{d\lambda}\right)^{2}-2g_{t\phi}\frac{dt}{d\lambda}\frac{d\phi}{d\lambda}-g_{\phi\phi}\left(\frac{d\phi}{d\lambda}\right)^{2}-g_{rr}\left(\frac{dr}{d\lambda}\right)^{2}=0,
\end{equation}
for photons in the $z=0$ plane. We then take the affine parameter along the photon trajectory to be the coordinate time $\lambda=t$, so that the impact parameter becomes 
\begin{equation}
    b=\frac{J}{E}=\frac{g_{t\varphi}+g_{\varphi\varphi}\frac{d\varphi}{d\lambda}}{-g_{tt}-g_{t\varphi}\frac{d\varphi}{d\lambda}}.
\end{equation}
This equation can be inverted to obtain 
\begin{equation}
    \frac{d\varphi}{d\lambda}=\frac{-bg_{tt}-g_{t\varphi}}{g_{\varphi\varphi}+bg_{t\varphi}}
\end{equation}
Since we are interested in the calculation of the deflection angle, that is $\varphi$ as a function of $r$, we eliminate the affine parameter in terms of the impact parameter and substitute in \eqref{epsilon},
\begin{equation}
    \left(\frac{dr}{d\varphi}\right)^{2}=\frac{(-g_{tt}g_{\varphi\varphi}+g_{t\varphi}^{2})(g_{\varphi\varphi}+2bg_{t\varphi}+b^{2}g_{tt})}{g_{rr}(bg_{tt}+g_{t\varphi})^{2}}.
\end{equation}
As derived in Sec. \ref{BG model}, the metric components for a rigidly rotating dust solution are $g_{tt}=-1$, $g_{t\varphi}=N$, $g_{rr}=e^{\mu}$, so that
\begin{equation}
    \left(\frac{dr}{d\varphi}\right)^{2}=\frac{r^{2}e^{-\mu}}{(N-b)^{2}}[r^{2}-(N-b)^{2}].
\end{equation}
Considering a photon coming towards the galaxy from very large distances compared to the galaxy radius, with an incident direction $\phi_{\infty}$, the solution may be determined by a quadrature,
\begin{equation}
\label{deflection_angle}
    \varphi(r)=\varphi_{\infty}+\int_{r}^{\infty}\frac{\abs{N(r)-b}e^{\frac{\mu}{2}}}{r^{2}\left[1-\frac{(N(r)-b)^{2}}{r^{2}}\right]^{\frac{1}{2}}}.
\end{equation}
If we denote $r_{m}$ the distance of minimum approach, then $\frac{dr}{d\varphi}|_{r_{m}}=0$ and we can write the impact parameter as
\begin{equation}
    b=N(r_{m})\pm r_{m}.
\end{equation}
Substituting this expression in \eqref{deflection_angle} we obtain
\begin{equation}
\label{deflection angle in function of rm}    \varphi(r)=\varphi_{\infty}+\int_{r}^{\infty}\frac{e^{\frac{\mu}{2}}\abs{N(r)-N(r_{m})\mp r_{m}}}{r^{2}\left[1-\frac{1}{r^{2}}(N(r)-N(r_{m})\mp r_{m})^{2}\right]^{1/2}}dr.
\end{equation}
The idea is now to use the expressions for $N(r)$ and $\mu(r)$ found in the previous sections in the asymptotic limit $r\gg R$. In particular, expanding \eqref{eq: simplified expression of N for the particular choice of spectral density} on the galactic plane ($z=0$) for $r\gg R$ we have
\begin{equation}
\label{N for r>>R}
    N(r)\simeq v_{0}(R-r_{0})+\frac{v_{0}(r_{0}^{2}-R^{2})}{2r}-\frac{v_{0}(r_{0}^{4}-R^{4})}{8r^{3}}\equiv N_{0}+\frac{c_{1}}{r}+\frac{c_{2}}{r^{3}}.
\end{equation}
In this limit, using \eqref{eq: explicit expression for mu}, the factor $e^{\frac{\mu}{2}}$ becomes
\begin{equation}
\label{e^mu/2 for r>>R}
    e^{\frac{\mu(r)}{2}}\simeq e^{\frac{\mu_{0}}{2}}\left\{\left[\frac{16r_{0}^{2}R^{2}}{(r_{0}+R)^{4}}\right]^{\frac{v_{0}^{2}}{8}}+\left[\frac{16r_{0}^{2}R^{2}}{(r_{0}+R)^{4}}\right]^{\frac{v_{0}^{2}}{8}}\frac{v_{0}^{2}}{64}(r_{0}^{4}+R^{4}-3r_{0}^{2}R^{2})\frac{1}{r^{4}}\right\}\equiv e^{\frac{\mu_{0}}{2}}\left(\alpha+\frac{\beta}{r^{4}}\right).
\end{equation}
Inserting \eqref{N for r>>R} and \eqref{e^mu/2 for r>>R} in equation \eqref{deflection angle in function of rm} for the deflection angle we obtain
\begin{align}
    \varphi(r)=\varphi_{\infty}+\int_{r}^{\infty}\frac{e^{\frac{\mu_{0}}{2}}\left(\alpha+\frac{\beta}{r^{4}}\right)\abs{\frac{c_{1}}{r_{m}r}+\frac{c_{2}}{r_{m}r^{3}}-\frac{c_{1}}{r_{m}^{2}}-\frac{c_{2}}{r_{m}^4}\mp 1}}{\left[1-\frac{r_{m}^{2}}{r^{2}}\left(\frac{c_{1}}{r_{m}r}+\frac{c_{2}}{r_{m}r^{3}}-\frac{c_{1}}{r_{m}^{2}}-\frac{c_{2}}{r_{m}^{4}}\mp 1\right)^{2}\right]^{\frac{1}{2}}}\frac{r_{m}dr}{r^{2}}.
\end{align}
We then proceed by expanding the integrand to the fourth order in $r^{-1}$ and $r_{m}^{-1}$, taking care of the fact that $r$ and $r_{m}$ are considered to be of the same order. The result of this procedure is 
\begin{equation}
    \varphi(r)-\varphi_{\infty}=\int_{r}^{\infty}\frac{e^{\frac{\mu_{0}}{2}}\left(\pm\alpha-\frac{\alpha c_{1}r}{r_{m}^{2}(r+r_{m})}\pm\frac{3\alpha c_{1}^{2}}{2r_{m}^{2}(r+r_{m})^{2}}-\frac{\alpha c_{2}(r^{2}+r_{m}r+r_{m}^{2})}{r_{m}^{4}r(r+r_{m})}\pm\frac{\beta}{r^{4}}\right)}{\left[\left(\frac{r}{r_{m}}\right)^{2}-1\right]^{1/2}}\frac{dr}{r}.
\end{equation}
The five integrals appearing in the above expression can now be computed analytically to give
\begin{align}
    \varphi(r)&=\varphi_{\infty}+e^{\frac{\mu_{0}}{2}}\left\{\pm\alpha\left(\frac{\pi}{2}-\arctan{\frac{\sqrt{r^{2}-r_{m}^{2}}}{r_{m}}}\right)-\frac{\alpha c_{1}}{r_{m}^{2}}\left(1-\frac{\sqrt{\frac{r}{r_{m}}-1}}{\sqrt{\frac{r}{r_{m}}+1}}\right)\right. \notag \\
    &\left. \pm\frac{3\alpha c_{1}^{2}}{2r_{m}^{4}}\left[\frac{\pi}{2}-\frac{3}{2}+\frac{1}{6}+\arcsec{\left(\frac{r}{r_{m}}\right)}-\frac{3}{2}\frac{\sqrt{\frac{r}{r_{m}}-1}}{\sqrt{\frac{r}{r_{m}}+1}}+\frac{1}{6}\frac{\left(\frac{r}{r_{m}}-1\right)^{3/2}}{\left(\frac{r}{r_{m}}+1\right)^{3/2}}\right]\right. \notag \\
    &\left.-\frac{\alpha c_{2}}{r_{m}^{4}}\left[2-\frac{r_{m}(2r+r_{m})}{r(r+r_{m})}\sqrt{\left(\frac{r}{r_{m}}\right)^2-1}\right]\right. \notag \\
    &\left. \pm\frac{\beta}{r_{m}^{4}}\left[\frac{3\pi}{16}+\frac{3}{8}\arctan{\left(\frac{\sqrt{r^{2}-r_{m}^{2}}}{r_{m}}\right)}+\frac{\sqrt{r^{2}-r_{m}^{2}}(3r_{m}r^{2}+2r_{m}^{3})}{8r^{4}}\right]\right\}.
\end{align}

\begin{figure}[H]
\centering
\includegraphics[scale=0.8]{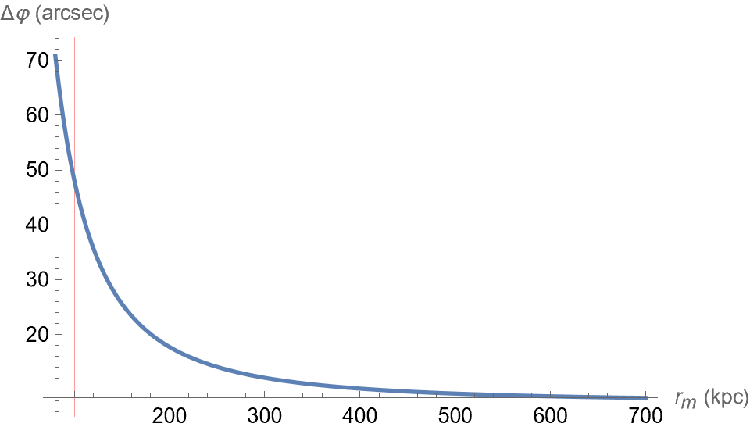}
\caption{Plot of $\Delta\varphi$ as a function of the distance of minimum approach $r_{m}$ for the values of the parameters reported above.}
\label{plot deflection angle}
\end{figure}

Now, to obtain the deflection of the orbit from a straight line it is important to keep in mind that the spacetime is not asymptotically Minkowski, due to the presence of the conical geometry. Therefore, the deflection angle is given by\footnote{For generality, in this expression we kept also the constant $\mu_0$. The latter must be set to $0$ if we want to eliminate the conical singularity.}
\begin{align}
\label{delta phi}
    \Delta\varphi&=2|\varphi(r_{m})-\varphi_{\infty}|-\lambda^{-1}e^{\frac{\mu_{0}}{2}}\pi  =2e^{\frac{\mu_{0}}{2}}\abs{\pm\alpha\frac{\pi}{2}-\frac{\alpha c_{1}}{r_{m}^{2}}\pm\frac{\alpha c_{1}^{2}}{r_{m}^{4}}\left(\frac{3\pi}{4}-2\right)-\frac{2\alpha c_{2}}{r_{m}^{4}}\pm\frac{\beta}{r_{m}^{4}}\frac{3\pi}{16}}-\lambda^{-1}e^{\frac{\mu_{0}}{2}}\pi.
\end{align}
To give a numerical estimate of the deflection angle in a Balasin-Grumiller galaxy model, we can now consider the typical parameters of a BG galaxy:
\begin{align}
    r_{0}=1 Kpc=3.09\times 10^{19} m, \qquad\  R=100 Kpc= 3.09\times 10^{21} m, \qquad\   v_{0}=200\frac{km}{s}=6.67\times 10^{-4}.
\end{align}
Substituting these values in the constants appearing in \eqref{N for r>>R} and \eqref{e^mu/2 for r>>R}, we obtain
\begin{align}
    &c_{1}=\frac{v_{0}}{2}(r_{0}^{2}-R^{2})=-3.18\times 10^{39}\, m^{2}, \\
    &c_{2}=-\frac{v_{0}}{8}(r_{0}^{4}-R^{4})=7.60\times 10^{81}\, m^{4}, \\
    &\alpha=\left[\frac{16 r_{0}^{2}R^{2}}{(r_{0}+R)^{4}}\right]^{\frac{v_{0}^{2}}{8}}=0.9995, \\
    &\beta=\left[\frac{16 r_{0}^{2}R^{2}}{(r_{0}+R)^{4}}\right]^{\frac{v_{0}^{2}}{8}}\frac{v_{0}^{2}}{64}(r_{0}^{4}+R^{4}-3r_{0}^{2}R^{2})=5.07\times 10^{78}\, m^{4}\\
    &e^{\frac{\mu_{0}}{2}}=0.295, \\
    &\lambda^{-1}=0.99946.
\end{align}
The plot of the deflection angle as a function of the distance of minimum approach is shown in Fig. \ref{plot deflection angle}.
This calculation predicts a behavior of the deflection angle that is consistent with that obtained in \cite{Galoppo2022}, in which the conical topology was neglected. This agreement is expected for the BG model, as the effect of the asymptotic conical topology (given by $\lambda$) is minimal for the chosen parameter values.
It is important to note that expression (\ref{delta phi}) obtained for the BG model includes a substantial contribution from the angular defect along the symmetry axis.
However, as previously discussed, the BG model assumes rigid rotation of the dust making it unsuitable for describing a whole galaxy.
For a dust configuration without conical singularity ($\mu_0=0$) all the effect would arise solely from the $\lambda$ factor which encodes the asymptotically conical structure. In this case there would be considerable differences in the deflection angles as measured by global and local observers. Specifically, the result \eqref{delta phi} reflects the perspective of a global observer measuring the deflection angle in a conical manifold, whereas a local observer, assuming an asymptotically Minkowski spacetime, would obtain significantly different values. \\
This highlights the importance of the conical geometry around the rotation axis (whether or not it arises from a conical singularity) in calculating the gravitational deflection of light within these general relativistic models.

\subsection{Propagation of a scalar field}
\label{quantum}
While the measure of the holonomy matrix and the gravitational lensing provide classical tests for the presence of conical singularities, quantum effects offer an additional perspective. The main reason is that quantum mechanics and quantum field theories are sensitive to global properties; therefore, the propagation of a quantized field on a conical manifold can provide deeper insights into the geometry of general relativistic models of galaxies. On a more general note, quantum observables complement the classical tests and offer a broader understanding of the non-Newtonian features in such models. 
\\
We are interested in the asymptotic behavior of the fields; and in particular, to the asymptotic particle states which are determined by the properties of the Riemann-flat and conical region.
\\
For this reason we will work with the conical geometry as an approximation of the true underlying geometry.
Let us consider the global and local observers introduced in Section \ref{holonomy method}. The global observer (GO) knows about the existence of an angular defect and describes the spacetime through a metric of the form
\begin{equation}
\label{conical metric}
    ds^{2}=-dt^{2}+dr^{2}+\frac{r^{2}}{\mu^{2}}d\theta^{2}+dz^{2},
\end{equation}
with $\theta\in [0,2\pi]$. On the other hand, the local observer (LO) has access only to a limited angular region and is induced to believe that the spacetime is also globally Minkowski,
\begin{equation}
\label{LO metric}
    ds^{2}=-dt^{2}+dr^{2}+r^{2}d\Tilde{\theta}^{2}+dz^{2},
\end{equation}
with $\Tilde{\theta}\in[0,2\pi]$. As already discussed the conical metric \eqref{conical metric} is only locally isomorphic to Minkowski since $\Tilde{\theta}=\frac{\theta}{\mu}$; therefore, the two observers use different background metrics when quantizing the fields. As we are going to demonstrate, this leads to an ambiguity in the definition of a common vacuum state and to a different interpretation of the field by the local observer. \\
As a working setting we consider a scalar field propagating in the $(2+1)-$dimensional spacetime described by the metric \eqref{conical metric} with $z=const.$
Within the physical lore given above, this is the reference frame of a global observer (GO) who knows about the global geometry of spacetime. The Klein-Gordon equation is,
\begin{equation}
    \frac{1}{\sqrt{-g}}\partial_{\mu}(\sqrt{-g}g^{\mu\nu}\partial_{\nu}\phi)-M^{2}\phi=0,
\end{equation}
where $\sqrt{-g}=\sqrt{-\det g_{\mu\nu}}$ and $M$ is the field mass. Adopting cylindrical coordinates we obtain
\begin{equation}
\label{KG equation cone}
    \left(-\partial_{t}^{2}+\frac{1}{r}\partial_{r}+\partial_{r}^{2}+\frac{\mu^{2}}{r^{2}}\partial_{\theta}^{2}-M^{2}\right)\phi=0,
\end{equation}
which can be solved assuming a separation of variables,
\begin{equation}
    \phi(t,r,\theta)=T(t)\psi(r,\theta).
\end{equation}
The resulting system for $T$ and $\psi$ reads
\begin{equation}
    \begin{cases}
        \partial_{t}^{2}T(t)= -(k^{2}+M^{2})T(t), \\
        \partial_{r}^{2}\psi(r,\theta)+\frac{1}{r}\partial_{r}\psi(r,\theta)+\frac{\mu^{2}}{r^{2}}\partial_{\theta}^{2}\psi(r,\theta)=-k^{2} \psi(r,\theta).
    \end{cases}
\end{equation}
A complete basis of solutions for the first equation is given by the plane waves
\begin{equation}
    T(t)\sim e^{\pm i\omega_{k}t},
\end{equation}
where $\omega_{k}=\sqrt{k^{2}+M^{2}}$. To solve the second we perform a further separation of variables,
\begin{equation}
    \psi(r,\theta)=R(r)\Theta(\theta),
\end{equation}
and obtain the system
\begin{equation}
    \begin{cases}
        \partial_{\theta}^{2}\Theta(\theta)=-m^{2}\Theta(\theta) \\
        \partial_{r}^{2}R(r)+\frac{1}{r}\partial_{r}R(r)+\left(k^{2}-\frac{\mu^{2}m^{2}}{r^{2}}\right)R(r)=0.
    \end{cases}
\end{equation}
The solution to the first equation is
\begin{equation}
    \Theta(\theta)\sim e^{im\theta},
\end{equation}
with $m\in\mathbb{Z}$ to ensure $2\pi$-periodicity. The remaining equation is solved by the combination
\begin{equation}
    R(r)= aJ_{\mu\abs{m}}(kr)+bY_{\mu\abs{m}}(kr),
\end{equation}
where $J_{\mu\abs{m}}(kr)$ and $Y_{\mu\abs{m}}(kr)$ are Bessel functions of first and second kind respectively. To guarantee regularity at the origin (ensuring essential self-adjointness) we take $b=0$. \\
Putting everything together, the positive frequency mode solutions to the Klein-Gordon equation are
\begin{equation}
\label{conical mode solutions}
    u_{km}(t,r,\theta)=J_{\mu\abs{m}}(kr)e^{im\theta}e^{-i\omega_{k}t},
\end{equation}
which we normailze with respect to the following scalar product 
\begin{equation}
    \label{scalar product}
    (\phi|\psi)=-i\int d^{2}x((\partial_{t}\phi^{*})\psi+\phi^{*}(\partial_{t}\psi)).
\end{equation}
The general solution to the Klein-Gordon equation can thus be written as
\begin{equation}
\label{conical field}
    \phi=\sum_{m\in\mathbb{Z}}\frac{1}{2\pi}\int_{0}^{\infty}\frac{dk}{2\omega_{k,m}}\mu k[a_{k,m}J_{\mu\abs{m}}e^{im\theta}e^{-i\omega_{k}t}+a^{\dagger}_{k,m}J_{\mu\abs{m}}e^{-im\theta}e^{i\omega_{k}t}].
\end{equation}
We now perform a canonical quantization of the field. The conjugate momentum is
\begin{align}
    \Pi(x)&=\frac{\partial\mathcal{L}}{\partial(\partial_{t}\phi)}=\partial_{t}\phi\\
    &=-\frac{i}{2\pi}\sum_{m\in\mathbb{Z}}\int_{0}^{\infty}\frac{dk}{2}\mu k[a_{k,m}J_{\mu\abs{m}}e^{im\theta}e^{-i\omega_{k}t}-a^{\dagger}_{k,m}J_{\mu\abs{m}}e^{-im\theta}e^{i\omega_{k}t}].
\end{align}
Imposing the equal time canonical commutation relations,
\begin{align}
    &[\phi(t,r,\theta),\phi(t,r',\theta')]=0, \\
    &[\Pi(t,r,\theta),\Pi(t,r',\theta')]=0, \\
    &[\phi(t,r,\theta),\Pi(t,r',\theta')]=\frac{i}{r}\delta(r-r')\delta(\theta-\theta'),
\end{align}
the commutation relations between the annihilation and creation operators follow. We exploit the scalar product \eqref{scalar product} to project the field on the normal modes \eqref{conical mode solutions}
\begin{align}
    &a_{k,m}=(u_{k,m}|\phi), \\
    &a^{\dagger}_{k,m}=-(u^{*}_{k,m}|\phi),
\end{align}
obtaining
\begin{align}
\label{commutation ladders}
    &[a_{k,m},a_{k',m'}]=0, \\
    &[a^{\dagger}_{k,m},a^{\dagger}_{k',m'}]=0, \\
    &[a_{k,m},a^{\dagger}_{k',m'}]=\frac{4\pi}{\mu}\frac{\omega_{k}}{k}\delta(k-k')\delta_{m,m'}.
\end{align}
The vacuum state $\ket{k=0,m=0}:=\ket{0_{G}}$ is defined as the state containing no particles according to all global inertial observers, i.e.,
\begin{equation}
    a_{k,m}\ket{0_{G}}=0 \quad \forall k\in{\mathbb{R^{+}}}, \forall m\in\mathbb{Z}.
\end{equation}
The particle states are then built by applying the creation operator to the vacuum state,
\begin{equation}
\label{N-particle state LO}    \ket{k_{1},m_{1};k_{2},m_{2};...;k_{N},m_{N}}=a^{\dagger}_{k_{1},m_{1}}a^{\dagger}_{k_{2},m_{2}}...a^{\dagger}_{k_{N},m_{N}}\ket{0_{G}}.
\end{equation}
As for the asymptotically local observers (LOs), they will write the metric as \eqref{LO metric} setting again $z=const.$ and assume that $\Tilde{\theta}$ has period $2\pi$. Keep in mind that the angular variable in the LO frame is of course connected to the one in the GO frame by $\Tilde{\theta}=\frac{\theta}{\mu}$ and has in fact period $\frac{2\pi}{\mu}$, but the LO is not aware of this. The LO thus describes the field through the flat normal modes 
\begin{equation}
\label{flat modes}
    v_{k,m}(t,r,\Tilde{\theta})=J_{\abs{m}}(kr)e^{im\Tilde{\theta}}e^{-i\omega_{k}t},
\end{equation}
so that 
\begin{equation}
\label{local field}
    \phi=\sum_{m\in\mathbb{Z}}\frac{1}{2\pi}\int_{0}^{\infty}\frac{dk}{2\omega_{k}} k[\tilde{a}_{k,m}J_{\abs{m}}e^{im\tilde{\theta}}e^{-i\omega_{k}t}+\tilde{a}^{\dagger}_{k,m}J_{\abs{m}}e^{-im\Tilde{\theta}}e^{i\omega_{k}t}].
\end{equation}
We can now apply the canonical quantization as for the GO case. In particular, we define the vacuum state $\ket{k=0,m=0}_{L}:=\ket{\tilde{0}_{L}}$ as the state that contains no particles as measured by all inertial local observers,
\begin{equation}
\label{local vacuum}
    \tilde{a}_{k,m}\ket{\tilde{0}_{L}}=0 \quad \forall k\in\mathbb{R^{+}}, \forall m\in\mathbb{Z}.
\end{equation}
Again, the particle states result from the application of the local creation operator  
\begin{equation}
    \ket{k_{1},m_{1};k_{2},m_{2};...;k_{N},m_{N}}_{L}=a^{\dagger}_{k_{1},m_{1}}a^{\dagger}_{k_{2},m_{2}}...a^{\dagger}_{k_{N},m_{N}}\ket{\tilde{0}_{L}}.
\end{equation}
It is important to stress that the flat normal modes \eqref{flat modes} form a complete basis only in Minkowski spacetime assuming $\Tilde{\theta}\in[0,2\pi)$. Therefore, only an observer confined in a limited region of spacetime would conclude that the field can be expanded as in \eqref{local field}, while a global observer would conclude that the correct expression is \eqref{conical field}. Nonetheless, the two observers would agree that the two expressions for the field coincide locally, since the conical metric \eqref{conical metric} is locally isomorphic to Minkowski. \\
We want now to understand the relation between the particle states defined by the two classes of observers. In order to study this aspect, we expand the annihilation operators of the LO in terms of those of the GO by computing the projection of the field \eqref{conical field} on the local normal modes \eqref{flat modes}.
\begin{align}
    \Tilde{a}_{k,n}&=(v_{k,n}|\phi) \\
    =&-i\int_{0}^{\infty}dr r \int_{a}^{a+2\pi}d\Tilde{\theta}((\partial_{t}v^{*}_{k,n})\phi-v^{*}_{k,n}\partial_{t}\phi)= \notag \\
    =&\sum_{m\in\mathbb{Z}}\int_{0}^{\infty}\frac{dk'}{2\omega_{k'}}\mu k'\int_{0}^{\infty}dr r J_{\abs{n}}(kr)J_{\mu\abs{m}}(k'r) \notag \\
    &\left[\int_{a}^{a+2\pi}\frac{d\Tilde{\theta}}{2\pi}e^{-in\Tilde{\theta}}e^{im\theta}(\omega_{k}+\omega_{k'})e^{i(\omega_{k}-\omega_{k'})t}a_{k',m}+\right. \notag \\
    &\left.+\int_{a}^{a+2\pi}\frac{d\Tilde{\theta}}{2\pi}e^{-in\Tilde{\theta}}e^{im\theta}(\omega_{k}-\omega_{k'})e^{i(\omega_{k}+\omega_{k'})t})a^{\dagger}_{k',m}\right].
    \label{eq: flat annihilation operator as a combination of the conical ladder operators}
\end{align}
The integrals in $\Tilde{\theta}$ can be easily performed if we remember that $\tilde{\theta}=\frac{\theta}{\mu}$,
\begin{equation}
    \int_{a}^{a+2\pi}\frac{d\Tilde{\theta}}{2\pi}e^{-in\Tilde{\theta}}e^{im\theta}=
    \int_{a}^{a+2\pi}\frac{d\Tilde{\theta}}{2\pi}e^{-i(n-\mu m)\Tilde{\theta}}=e^{-in(a+\pi)}e^{i\mu m(a+\pi)}\frac{\sin{[\pi(n-\mu m)]}}{\pi(n-\mu m)}.
\end{equation}
We then obtain,
\begin{align}
\label{local annihilation operators}
    \Tilde{a}_{k,n}&=e^{-in(a+\pi)}\sum_{m\in\mathbb{Z}}\int_{0}^{\infty}\frac{dk'}{2\omega_{k'}}\mu k'\int_{0}^{\infty}dr r J_{\abs{n}}(kr)J_{\mu\abs{m}}(k'r) \notag \\
    &\left(e^{i\mu m(a+\pi)}\frac{\sin{[\pi(n-\mu m)]}}{\pi(n-\mu m)}(\omega_{k}+\omega_{k'})e^{i(\omega_{k}-\omega_{k'})t}a_{k',m}+\right. \notag \\
    &\left.e^{-i\mu m(a+\pi)}\frac{\sin{[\pi(n+\mu m)]}}{\pi(n+\mu m)}(\omega_{k}-\omega_{k'})e^{i(\omega_{k}+\omega_{k'})t}a^{\dagger}_{k',m}\right).
\end{align}
A possible way to calculate the integral over the radial variable is discussed in Appendix B. Here we report the result, 
\begin{align}
\label{integral Bessel}
    I_{m,n}(k',k)&=\int_{0}^{\infty}rJ_{\mu{|m|}}(k'r)J_{{|n|}}(kr)dr\cr
     =&\frac{1}{k}\cos{\left[\frac{(\mu |m|-|n|)}{2}\pi\right]}\delta(k-k')+\frac 1{k'-k}\frac 1\pi \sin \Big(\frac\pi2 (\mu |m|-|n|)\Big) \left( \frac {2}{k+k'} \Phi_{mn}(k,k')-\frac {1}{\sqrt {kk'}} \right),
\end{align}
where
\begin{align}
    \Phi_{mn}(k,k')=& \theta(k-k') \left(\frac{k^{\prime}}{k}\right)^{\mu|m|}  \frac {\Gamma(\frac \mu2 |m|-\frac {|n|}2+1)\Gamma(\frac \mu2 |m|+\frac {|n|}2+1)}{\Gamma(\mu |m|+1)} \times \cr 
    &\times{ }_2 F_1\left(\frac \mu2 |m|+\frac {|n|}2, \frac \mu2 |m|-\frac {|n|}2 ; \mu |m|+1 ; \frac{k^{\prime 2}}{k^2}\right)\cr
    -&\theta(k'-k) \left(\frac{k}{k'}\right)^{|n|}  \frac {\Gamma(\frac {|n|}2-\frac \mu2 |m|+1)\Gamma(\frac \mu2 |m|+\frac {|n|}2+1)}{\Gamma(|n|+1)} \times \cr &\times { }_2 F_1\left(\frac \mu2 |m|+\frac {|n|}2, \frac {|n|}2-\frac \mu2 |m| ; |n|+1 ; \frac{k^{2}}{k'^2}\right),
\end{align}
and
\begin{align}
    \Phi_{mn}(k,k)=1.
\end{align}
In fact, $\Phi_{mn}(k,k')$ is a function of $k/k'$ only and we can write
\begin{align}
    \Phi_{mn}(k,k')\equiv \psi^{(\mu)}_{mn}(k'/k).
\end{align}
From expression \eqref{local annihilation operators} we see that the vacuum state defined by global observers is not empty from the point of view of local ones.
\\Indeed, if we apply the annihilation operator defined by a LO to the GO vacuum we obtain
\begin{align}
    \tilde{a}_{k,n}\ket{0_{G}}=&e^{-in(a+\pi)}\sum_{m\in\mathbb{Z}}\int_{0}^{\infty}\frac{dk'}{2\omega_{k'}}\mu k'I_{m,n}(k',k) \notag \\
    &\times e^{-i\mu m(a+\pi)}\frac{\sin{[(n-\mu m)\pi]}}{\pi(n-\mu m)}[(\omega_{k}-\omega_{k'})e^{i(\omega_{k}+\omega_{k'})t}]\ket{k',m}_{G}\neq 0,
\end{align}
in contrast to \eqref{local vacuum}. Therefore, the vacuum states built by the two classes of observers, and thus their entire Fock spaces, are different. 
Notice that the term $\delta(k-k')$ in $I_{nm}$ does not contribute to the integral while the remaining term scales like $k^{-\frac 32}$, so that the integrand is of order $\sim k^{-\frac 12} e^{i(\omega_{k}+\omega_{k'})t} \ket{k',m}_{G}$. This indicates that each state $\ket{k',m}_{G}$ is weighted by a non square integrable weight $h^{-\frac 12}$; therefore, the states $\tilde{a}_{k,n}\ket{0_{G}}$ and thus $\ket{0_{G}}$ are not in the Fock space of the local observer, meaning that the two theories cannot be related by a unitary map. The reason is most probably due to the wrong analysis made by the LO: the erroneous assumption on the periodicity of the angular variable makes the Hamiltonian operator non-self-adjoint on the global manifold. On the other hand, it is interesting to notice that we have considered only a very specific theory of a massive Klein-Gordon quantum field on a conical spacetime. 
A complete study of the problem has been considered in \cite{Kay}, where it was proved that there exists a whole one parameter family of self-adjoint Laplacians on the cone, leading to infinitely many variations of the QFT presented here. In particular, the version we discussed corresponds to the one regular at the conical singularity.
The point is that the different QFTs depend on the specific boundary conditions imposed at the singularity, which in turn determine the allowed behaviors of the elements of the Hilbert space. But here we are considering ACMs only, so it makes no sense to choose any specific details at the conifold singularity (being it not even present).
\\
Indeed, if we think of applying the dust model to the description of disc galaxies, the conical structure is supposed to be present only asymptotically away from the rotation axis since the solution would be modified in the bulge to take pressure into account. 
Now, since the boundary conditions on the axis are expected to affect only the UV spectrum, it is reasonable to introduce a cutoff in the momentum $k$. The latter is determined by the scales at which particles become sensitive to the local geometry at the singularity, whose size is the Compton wavelength or a fraction of it. Hence, let us introduce a momentum cutoff $k=M$ and proceed with the analysis. \\
From the above expression, we can compute the number of particles with quantum numbers $n$ and $k$ (in the invariant volume element $\frac {kdk}{4\pi \omega_k}$) seen by the local observer in the state $\ket{0_{G}}$. This is given by
\begin{align}
  dN(k,n)=&\|\tilde{a}_{k,n}\ket{0_{G}}\|^2 \frac {kdk}{4\pi \omega_k} \cr
  =& \left[\sum_{m\in\mathbb{Z}}\int_{0}^{M}\frac{dk'}{2\omega_{k'}}\mu k'I_{m,n}(k',k)^2 \frac{\sin^2{[(n-\mu m)\pi]}}{\pi^2(n-\mu m)^2}(\omega_{k}-\omega_{k'})^2\right] \frac {kdk}{2\omega_k}\cr
  = \Bigg[\sum_{m\in\mathbb{Z}}&\int_{0}^{M}\frac{dk'}{2\omega_{k'}}\mu k' 
  \frac{\sin^2{[(n-\mu m)\pi]}}{\pi^2(n-\mu m)^2}
  \frac {(\omega_{k}-\omega_{k'})^2}{(k'-k)^2} \frac{\sin^2 \Big(\frac\pi2 (\mu |m|-|n|)\Big)}{\pi^2}\cr
  &\times \left( \frac {2}{k+k'} \Phi_{mn}(k,k')-\frac {1}{\sqrt {kk'}} \right)^2\Bigg]\frac {kdk}{2\omega_k},
\end{align}
and can be rewritten in the form
\begin{align}
  dN(k,n)=&\Bigg[\mu\sum_{m\in\mathbb{Z}} 
  \frac{\sin^2{[(n-\mu m)\pi]}}{2\pi^2(n-\mu m)^2}
   \frac{\sin^2 \Big(\frac\pi2 (\mu |m|-|n|)\Big)}{\pi^2} \Lambda_{m,n}(\mu,M/k)\Bigg]\frac {dk}{2\omega_k},
\end{align}
where
\begin{align}
    \Lambda_{m,n}(\mu,x)=\int_{0}^{x}\frac {zdz}{\sqrt{z^2+x^2}}\frac {(1+z)^2}{(\sqrt{1+x^2}+\sqrt{z^2+x^2})^2} \left( \frac {2}{1+z} \Psi^{(\mu)}_{mn}(z)-\frac {1}{\sqrt {z}} \right)^2.
\end{align}
These expressions are intended for $k<M$, that is $x\geq 1$.

For $\mu=1+\epsilon$, we see that $dn(k,n)$ vanishes for $\epsilon=0$, while for $|\epsilon|\ll 1$ we have 
\begin{align}
    \frac{\sin^2{[(n-\mu m)\pi]}}{2\pi^2(n-\mu m)^2} \frac{\sin^2 \Big(\frac\pi2 (\mu |m|-|n|)\Big)}{\pi^2}\sim k_{mn}\epsilon^2,
\end{align}
where
\begin{align}
   k_{mn}= \begin{cases}
       \frac {n^2}8 & \mbox{ if }\quad m=n, \\
       \frac {(n+1+2l)^2}{2\pi^2(2l+1)^2} & \mbox{ if }\quad m=n+2l+1,\quad l\in \mathbb Z, \\
       0 & \mbox{ if }\quad m=n+2l,\quad l\in \mathbb Z-\{0\}.
   \end{cases}
\end{align}
Hence, we can write 
\begin{align}
    dN(k,n)=\epsilon^2 \Sigma_n(M/k) \frac {dk}{2\omega_k},
\end{align}
with 
\begin{align}
   \Sigma_n (x)=\frac {n^2}8 \Lambda_{nn}(1,x)+\sum_{l\in\mathbb Z} \frac {(n+1+2l)^2}{2\pi^2(2l+1)^2} \Lambda_{(n+2l+1),n}(1,x).
\end{align}
Therefore, while for the GO the spacetime is empty with a conical (locally Riemann-flat) geometry, by assuming a globally Minkowskian spacetime the LO interprets the vacuum as filled of scalar particles, with a more or less complicated spectral distribution. Of course, this comes with a little puzzle for them, since spacetime would be expected to be curved by such distribution of particles.
\\
We can thus conclude that the presence of a conical singularity has a direct influence on the Fock space of a quantum field theory. As a consequence, global and local observers could realize that they are giving two different descriptions of the same spacetime by sending to each other quanta of a scalar field and comparing the outcomes.
\\
These effects provide a distinct signature of the non-Newtonian nature of the galaxy models, complementing the classical tests explored in previous sections. The interplay between classical and quantum observables in the presence of conical singularities reinforces the need for a fully relativistic treatment of the dynamics of extended objects and, eventually, opens the door to new observational tests that could further distinguish general relativity from its Newtonian approximation in astrophysical contexts.

\section{Conclusions}
\label{conclusions}
In this work we investigated the effects of the asymptotically conical topology in stationary and axisymmetric configurations of self-gravitating dust.
\\By restricting to the $z=0$ plane and by assuming that the metric becomes static and flat far away from the rotation axis, we identified the presence of an angular defect in this asymptotic region. Geometrically, the defect arises because the conformal factor multiplying the metric of the $2$-spaces orthogonal to the Killing vectors does not vanish in the limit $r\to\infty$.
This feature defines what we call Asymptotically Conical Minkowskian (ACM) spacetimes, which retain a locally flat but globally conical structure unlike the standard Minkowski space. When applied to model rotating discs of dust, as could be the case of galaxies outside the bulge and close to the galactic plane, this unique topology generates non-Newtonian features with potentially observable effects. If confirmed experimentally, the latter would definitely prove the need for a fully general relativistic description of galaxies.
\\
As a general rule, the conical topology can only be revealed by means of global observables. The first and very direct possibility we addressed is the detection of a nontrivial asymptotic holonomy. Specifically, we considered an observer parallel-transporting a test vector in a closed path around the symmetry axis: in this case, the conical defect results in a nontrivial rotation of the transported vector and can in principle be detected, say, by using a gyroscope. In the physical context of a galaxy this is of course an ideal experiment, because a satellite cannot complete a full outer circle in a reasonable time lapse. Nonetheless, it allowed us to define two reference observers: a \say{global} one, who has experienced a full round trip, and a \say{local} one, who has access to a limited region of spacetime only. In this respect we considered that, without further proof, a local observer would always extrapolate from the locally flat geometry and assume a globally Minkowski background.\\ 
As a more realistic classical test we considered the gravitational lensing in the outer region of the Balasin-Grumiller model by computing perturbatively the deflection angle. In this case, the presence of a conical singularity is essential to provide an angle which is compatible with the observations; however, in more refined dust configurations the asymptotically conical topology can be enough to explain such observable effect even without a conical singularity.\\
As a last test we considered the propagation of a scalar quantum field, which is intrinsically a nonlocal entity. We performed a canonical quantization in the shoes of the global observer, who is aware of the conical topology, and in those of the local observer, who makes the erroneous extrapolation that the background is globally Minkowski. We showed explicitly that the resulting Fock spaces cannot be mapped into one another, since they are physically different; and we discussed how, as a result, the local observer ends up mistaking the global vacuum as populated by a distribution of scalar particles. 
The bottom line is that an erroneous assumption on the global topology can have a considerable impact on the matter content perceived by a local observer. \\
In the case of galaxies, a nontrivial global topology is a pure general relativistic feature; hence, the tests discussed in this work support the need of general relativistic models for such extended rotating sources.

\vspace{1.5cm}
\section*{Acknowledgments}
M. F. and F. S. wish to thank O. F. Piattella, A. Massidda and G. Bianchi for stimulating discussions. 
M. F. is deeply thankful to R. Peron for the hospitality and the opportunity to present this work at IAPS-INAF in Roma.

\appendix
\section{Appendix A}
\label{Appendix A}
We report the expressions of the Christoffel symbols of the rigidly rotating dust metric \eqref{rigidly rotating dust metric}, together with their expansion for $r\gg R$ that are used in Sec.\ \ref{holonomy method}.
\begin{align}
    \Gamma^{t}_{\alpha\beta}&=
     \begin{pmatrix}
     0 & 0 & \frac{1}{2}\frac{N}{r^{2}}N_{,r} & \frac{1}{2}\frac{N}{r^{2}}N_{,z} \\
     0 & 0 & -\frac{1}{2}\frac{N^{2}}{r^{2}}N_{,r}-\frac{1}{2}N_{,r}+\frac{N}{r} & -\frac{1}{2}\frac{N^{2}}{r^{2}}N_{,z}-\frac{1}{2}N_{,z} \\
     \frac{1}{2}\frac{N}{r^{2}}N_{,r} & -\frac{1}{2}\frac{N^{2}}{r^{2}}N_{,r}-\frac{1}{2}N_{,r}+\frac{N}{r} & 0 & 0 \\
     \frac{1}{2}\frac{N}{r^{2}}N_{,z} & -\frac{1}{2}\frac{N^{2}}{r^{2}}N_{,z}-\frac{1}{2}N_{,z} & 0 & 0
     \end{pmatrix},
\end{align}
\begin{align}
    \Gamma^{\phi}_{\alpha\beta}&=
    \begin{pmatrix}
     0 & 0 & \frac{1}{2}\frac{1}{r^{2}}N_{,r} & \frac{1}{2}\frac{1}{r^{2}}N_{,z} \\
     0 & 0 & -\frac{1}{2}\frac{N}{r^{2}}N_{,r}+\frac{1}{r} & -\frac{1}{2}\frac{N}{r^{2}}N_{,z} \\
     \frac{1}{2}\frac{1}{r^{2}}N_{,r} & -\frac{1}{2}\frac{N}{r^{2}}N_{,r}+\frac{1}{r} & 0 & 0 \\
     \frac{1}{2}\frac{1}{r^{2}}N_{,z} & -\frac{1}{2}\frac{N}{r^{2}}N_{,z} & 0 & 0
    \end{pmatrix},
\end{align}
\begin{equation}
    \Gamma^{r}_{\alpha\beta}=
    \begin{pmatrix}
    0 & -\frac{1}{2}e^{-\mu}N_{,r} & 0 & 0 \\
    -\frac{1}{2}e^{-\mu}N_{,r} & -re^{-\mu}+e^{-\mu}NN_{,r} & 0 & 0 \\
    0 & 0 & \frac{\mu_{,r}}{2} & \frac{\mu_{,z}}{2} \\
    0 & 0 & \frac{\mu_{,z}}{2} & -\frac{\mu_{,r}}{2}
    \end{pmatrix},
\end{equation}
\begin{equation}
    \Gamma^{z}_{\alpha\beta}=
    \begin{pmatrix}
    0 & -\frac{1}{2}e^{-\mu}N_{,z} & 0 & 0 \\
    -\frac{1}{2}e^{-\mu}N_{,z} & e^{-\mu}NN_{,z} & 0 & 0 \\
    0 & 0 & -\frac{\mu_{,z}}{2} & \frac{\mu_{,r}}{2} \\
    0 & 0 & \frac{\mu_{,r}}{2} & \frac{\mu_{,z}}{2}
    \end{pmatrix}.
\end{equation}
The expansions of the Christoffel symbols used for the computation of the holonomy matrix in Sec.\ \ref{holonomy method} for $r\gg r_{0}$ on the galactic plane $z=0$ are 
\begin{equation}
\begin{aligned}
    &\christoffel{t}{\phi}{z} = \christoffel{\phi}{\phi}{z} = \christoffel{z}{t}{\phi} = \christoffel{z}{\phi}{\phi} = 0,\\  
    &\christoffel{t}{\phi}{r} \simeq \frac{V_0}{r}(R-r_0) - \frac{V_0}{2r^2}(R^2 - r_0^2 + (R-r_0)^2),\\
    &\christoffel{\phi}{\phi}{r} \simeq \frac{1}{r},\\
    &\christoffel{r}{t}{\phi} \simeq -\frac{V_0}{4r^2}e^{-\mu}(R^2-r_0^2),\\
    &\christoffel{r}{\phi}{\phi} \simeq e^{-\mu}\left(\frac{V_0}{2r^2}(R-r_0)(R^2-r_0^2) - r\right).
\end{aligned}
\end{equation}

\section{Appendix B}
\label{Appendix B}
We report the explicit calculation of the integral \eqref{integral Bessel},
\begin{equation}
    \int_{0}^{\infty}rJ_{\mu\abs{m}}(k'r)J_{\abs{n}}(kr)dr.
\end{equation}
We start by noticing that this integral is not convergent; therefore, it has to be understood as a distribution by considering the weak-limit 
\begin{equation}
    I_{mn}(k,k')=w-\lim_{R\to\infty}\int_{0}^{R}rJ_{\mu\abs{m}}(k'r)J_{\abs{n}}(kr)dr.
\end{equation}
It is convenient to rewrite the above integral exploiting the fact that the Bessel functions of the first kind form a complete set of solutions for the following differential equation
\begin{equation}
    \frac{d^{2}}{dr^{2}}J_{\nu}(kr)+\frac{1}{r}\frac{d}{dr}J_{\nu}(kr)+\left(k^{2}-\frac{\nu^{2}}{r^{2}}\right)J_{\nu}(kr)=0.
\end{equation}
Therefore, defining $f=J_{\mu\abs{m}}(k'r)$ and $g=J_{\abs{n}}(kr)$ we have 
\begin{equation}
\label{equation f}
    f''+\frac{1}{r}f'+\left(k'^{2}-\frac{\mu^{2}m^{2}}{r^{2}}\right)f=0,
\end{equation}
\begin{equation}
\label{equation g}
    g''+\frac{1}{r}g'+\left(k^{2}-\frac{n^{2}}{r^{2}}\right)f=0,
\end{equation}
where the prime indicates derivation with respect to $r$.
Multiplying the first by $g$ and the second by $f$ and then taking the difference yields
\begin{equation}
    \frac{d}{dr}(rgf'-rfg')+r(k'^{2}-k^{2})fg-\frac{\mu^{2}m^{2}-n^{2}}{r}fg=0.
\end{equation}
Integrating over $r$ we finally obtain
\begin{align}
\label{integral split}
    I_{mn}(k',k)=&\frac{\mu^{2}m^{2}-n^{2}}{k'^{2}-k^{2}}\int_{0}^{\infty}dr\frac{J_{\mu\abs{m}}(k'r)J_{\abs{n}}(kr)}{r} \notag \\
    &-w\lim_{R\to\infty}\frac{R}{k'^{2}-k^{2}}(k'J'_{\mu\abs{m}}(k'R)J_{\abs{n}}(kR)-kJ_{\mu\abs{m}}(k'R)J'_{\abs{n}}(kR)).
\end{align}
For $|m|+|n|\neq 0$, the first integral is convergent and it has been calculated in \cite{Watson} (chapter 13.4, formula (2)). The result is 

\begin{equation}
\label{integral watson}
\begin{aligned}
J_{mn}:=&\int_0^{\infty} d r \frac{J_{\mu|m|}\left(k^{\prime} r\right) J_{|n|}(k r)}{r} \cr 
=& \frac{\theta\left(k^{\prime}-k\right)}{2}\left(\frac{k}{k^{\prime}}\right)^{|n|}\frac{\Gamma(\frac \mu2 |m|+\frac {|n|}2)}{\Gamma(|n|+1) \Gamma(1+\frac \mu2 |m|-\frac {|n|}2)}{ }_2 F_1\left(\frac \mu2 |m|+\frac {|n|}2, \frac {|n|}2-\frac \mu2 |m| ; |n|+1 ; \frac{k^2}{k^{\prime 2}}\right) \cr
 &+\frac{\theta\left(k-k^{\prime}\right)}{2}\left(\frac{k^{\prime}}{k}\right)^{\mu|m|}\frac{\Gamma(\frac \mu2 |m|+\frac {|n|}2)}{\Gamma(\mu |m|+1) \Gamma(1-\frac \mu2 |m|+\frac {|n|}2)}{ }_2 F_1\left(\frac \mu2 |m|+\frac {|n|}2, \frac \mu2 |m|-\frac {|n|}2 ; \mu |m|+1 ; \frac{k^{\prime 2}}{k^2}\right).
\end{aligned}
\end{equation}
If we use $\Gamma(z)\Gamma(1-z)=\pi/\sin(\pi z)$ and define the function
\begin{align}
    \Phi_{mn}(k,k')=& \theta(k-k') \left(\frac{k^{\prime}}{k}\right)^{\mu|m|}  \frac {\Gamma(\frac \mu2 |m|-\frac {|n|}2+1)\Gamma(\frac \mu2 |m|+\frac {|n|}2+1)}{\Gamma(\mu |m|+1)} \times \cr 
    &\times{ }_2 F_1\left(\frac \mu2 |m|+\frac {|n|}2, \frac \mu2 |m|-\frac {|n|}2 ; \mu |m|+1 ; \frac{k^{\prime 2}}{k^2}\right)\cr
    -&\theta(k'-k)\left(\frac{k}{k'}\right)^{|n|}  \frac {\Gamma(\frac {|n|}2-\frac \mu2 |m|+1)\Gamma(\frac \mu2 |m|+\frac {|n|}2+1)}{\Gamma(|n|+1)} \times \cr &\times { }_2 F_1\left(\frac \mu2 |m|+\frac {|n|}2, \frac {|n|}2-\frac \mu2 |m| ; |n|+1 ; \frac{k^{2}}{k'^2}\right), \label{phimn}
\end{align}
then we can write
\begin{align}
    \frac {\mu^2 m^2-n^2}{k'^2-k^2} J_{mn}=\frac 2{k'^2-k^2} \frac 1\pi \sin (\frac \mu2 |m|-\frac {|n|}2) \Phi_{mn}(k,k').
\end{align}
It is useful to notice that
\begin{align}
    \Phi_{mn}(k,k)=1 \label{phikk}
\end{align}
is independent from $k$.  \\
Therefore, we are left with the calculation of the weak-limit in Eq.\ \eqref{integral split}, which can be performed remembering that
\begin{equation}
    J'_{\nu}(z)=-J_{\nu+1}(z)+\frac{\nu}{z}J_{\nu}(z),
\end{equation}
together with the asymptotic expansion of the Bessel functions
\begin{equation}
    J_{\nu}(z)\sim \sqrt{\frac{2}{\pi z}}\cos{\left(z-\frac{\nu}{2}\pi-\frac{\pi}{4}\right)}, \quad z\to\infty, \quad \abs{arg(z)}<\pi.
\end{equation}
After some simple algebra involving trigonometric identities, the limit can be recast in the form
\begin{align}
\label{weak_limit_long}
   w-\lim_{R\to\infty}&\left\{\cos{[(k'+k)R]}\cos{\left[\frac{(\mu\abs{m}+\abs{n})}{2}\pi\right]}\frac{1}{(k'+k)\pi\sqrt{kk'}} \right. \notag \\
   &\left. +\sin{[(k'+k)R]}\sin{\left[\frac{(\mu\abs{m}+\abs{n})}{2}\pi\right]}\frac{1}{(k'+k)\pi\sqrt{kk'}} \right. \notag \\
   &\left. -\sin{[(k'-k)R]}\cos{\left[\frac{(\mu\abs{m}-\abs{n})}{2}\pi\right]}\frac{1}{(k'-k)\pi\sqrt{kk'}} \right. \notag \\
   &\left. +\cos{[(k'-k)R]}\sin{\left[\frac{(\mu\abs{m}-\abs{n})}{2}\pi\right]}\frac{1}{(k'-k)\pi\sqrt{kk'}} \right. \notag \\
   &\left. +\frac{\mu\abs{m}-\abs{n}}{R(k'^{2}-k^{2})\pi\sqrt{kk'}}\left(\cos{[(k'-k)R]}\cos{\left[\frac{(\mu\abs{m}-\abs{n})}{2}\pi\right]}+\sin{[(k'-k)R]}\sin{\left[\frac{(\mu\abs{m}-\abs{n})}{2}\pi\right]} \right. \right. \notag \\
   &\left. \left.+\sin{[(k'+k)R]}\cos{\left[\frac{(\mu\abs{m}+\abs{n})}{2}\pi\right]}-\cos{[(k'+k)R]}\sin{\left[\frac{(\mu\abs{m}+\abs{n})}{2}\pi\right]}\right)
   \right\}.
\end{align}
The first two terms vanish because of the Riemann-Lebesgue theorem, while the last two are zero due to the presence of the $R$ factor at the denominator. The third term is 
\begin{equation}
\label{weak_limit_sin}
    w-\lim_{R\to\infty}\frac{\sin{[(k-k')R]}}{\pi\sqrt{kk'}(k-k')}=\frac{1}{k}\delta(k-k').
\end{equation}
Finally, the fourth term is easily calculated using the identity 
\begin{equation}
    \cos{[(k'-k)R]}=1-2\sin^{2}{\left[\frac{(k'-k)}{2}R\right]}.
\end{equation}
Putting everything together we obtain 
\begin{align}
\label{weak limit}
    w-\lim_{R\to\infty}\frac{R}{k'^{2}-k^{2}}(k'J_{\abs{n}}(kR)J'_{\mu\abs{m}}(k'R)-kJ_{\mu\abs{m}}(k'R)J'_{\abs{n}}(kR))=&-\frac{1}{k}\cos{\left[\frac{(\mu\abs{m}-\abs{n})}{2}\pi\right]}\delta(k-k') \notag \\
    &+\frac{1}{\pi(k'-k)\sqrt{kk'}}\sin{\left[\frac{(\mu\abs{m}-\abs{n})}{2}\pi\right]}.
\end{align}
Summing up things and using Eq.\ \eqref{phikk}, we can rewrite Eq.\ \eqref{integral split} as
\begin{align}
    I_{mn}(k,k')=\frac{1}{k}\cos{\left[\frac{(\mu |m|-|n|)}{2}\pi\right]}\delta(k-k')+\frac 1{k'-k}\frac 1\pi \sin (\frac \mu2 |m|-\frac {|n|}2) \left( \frac {2}{k+k'} \Phi_{mn}(k,k')-\frac {1}{\sqrt {kk'}} \right). \label{Imn}
\end{align}
The case $m=n=0$, the integral $J_{00}$ is multiplied by a zero factor and disappears, and the final calculation gives the same expression above with $m=n=0$, which is 
\begin{align}
    I_{00}(k,k')=\frac{1}{k} \delta(k-k').
\end{align}
We finally observe that in Eq.\ \eqref{Imn} only the delta factor is singular, while the second term in the sum for $k\sim k'$ behaves as
\begin{align}
    \frac {1}{k'-k} \left(\frac {2}{k+k'} -\frac 1{\sqrt {kk'}}\right)= \frac {1}{\sqrt{k'}+\sqrt k}\frac {\sqrt{k}-\sqrt {k'}}{(k+k')\sqrt {kk'}}\sim 0.
\end{align}

\bibliographystyle{unsrt}

\bibliography{bib.bib}

\end{document}